# Improved *Ab Initio* Molecular Dynamics by Minimally Biasing with Experimental Data

Andrew D. White,[1,2] Chris Knight,[3] Glen M. Hocky,[1] Gregory A. Voth[1]

[1]*Department of Chemistry, James Franck Institute, and Institute for Biophysical Dynamics, The University of Chicago, 5735 S Ellis Ave., Chicago, Illinois 60637, USA*

[2]*Department of Chemical Engineering, University of Rochester, Rochester, New York 14627, USA*

[3]*Leadership Computing Facility, Argonne National Laboratory, 9700 South Cass Avenue, Argonne, Illinois 60439, USA*

**Abstract**

Accounting for electrons and nuclei simultaneously is a powerful capability of *ab initio* molecular dynamics (AIMD). However, AIMD is often unable to accurately reproduce properties of systems such as water due to inaccuracies in the underlying electronic density functionals. This shortcoming is often addressed by added empirical corrections and/or increasing the simulation temperature. We present here a maximum-entropy approach to directly incorporate limited experimental data via a minimal bias. Biased AIMD simulations of water and an excess proton in water are shown to give significantly improved properties both for observables which were biased to match experimental data and for unbiased observables. This approach also yields new physical insight into inaccuracies in the underlying density functional theory as utilized in the unbiased AIMD.

*Ab initio* molecular dynamics (AIMD) is a class of simulation technique in which the forces governing motion of the nuclei are computed explicitly "on the fly" via quantum mechanical calculations of the electronic state of the system.[1] The principal advantages of AIMD over empirical classical molecular dynamics are that the electronic structure can polarize in response to the motion of the nuclei and that chemical bonds can break and form, allowing chemistry to occur in a dynamical simulation. The necessary complexity of accounting for electrons once meant AIMD was too slow to be of practical use for many applications, especially those of biomolecular relevance. This is no longer the case due to advances in AIMD algorithms and it is now fast enough to model meaningful aqueous systems such as hydrated excess proton diffusion in water.[2-4]

Despite these advances in computational speed, AIMD has had difficulty in accurately describing the most basic and fundamental of condensed phase systems: liquid water. The approximate density functional theory (DFT) commonly used in AIMD leads to an over-structuring of the AIMD water. Such simulated water at room temperature behaves as a glass, with a self-diffusion coefficient often as much as two orders of magnitude lower than the experimental value.[4-9] It has a melting point well above the correct one,[10-11] so that AIMD water simulated at 300K by these functionals is supercooled. The deficiencies of AIMD water structure and dynamics also influence excess proton diffusion in water, because excess protons can diffuse via hopping from molecule to molecule (the Grotthuss mechanism), breaking and forming chemical bonds.[12-14] Thus, proton transport, acid-base reactions, and the fundamental structure and dynamics of water are often difficult to properly describe by AIMD, not due to computational inefficiency but instead due to inaccuracies in the underlying DFT approximation.[15-16] There are a number of recent papers that have attempted to overcome this problem by using increasingly more complex exchange-correlation functionals,[17] with empirical corrections, or explicitly fitting the functional for water.[18] Nevertheless, the issues of an often over-structured AIMD liquid water and its low self-diffusion and glassy behavior remain. AIMD simulation of other molecular liquids may also be problematic due to the same or other inaccuracies in the underlying functionals,[19] but less focus has been devoted to such systems compared to water.

In the present work, we propose and demonstrate a fundamentally different approach to AIMD simulation. Instead of searching for a modification of the underlying DFT so that the AIMD simulation will accurately capture enough of the physics to match the experimental data, we adopt a method called experiment directed simulation[20] (EDS) to *directly* incorporate a minimal bias from a limited set of experimental data into the simulation. This is accomplished through the introduction of an auxiliary coupled "knowledge field" $V(\mathbf{r})$ which contains information on a minimal set of experimental observables. The system potential $U(\mathbf{r})$ becomes $U(\mathbf{r}) + V(\mathbf{r})$. EDS finds the unique and minimal change to a simulation Hamiltonian that produces an ensemble of configurations consistent with input data that would come from a target "exact" system. This latter feature of EDS has recently been mathematically proven within the context of developing an improved "bottom-up" coarse-grained model from an underlying imperfect atomistic potential function.[21] Moreover, it was shown that another form of biased MD, i.e., that with added harmonic constraints to the experimental data, will not minimize the relative entropy of the biased system with respect to the target exact system[21]. EDS is part of a growing body of work where molecular models are minimally modified to match experimental data, thus arriving at a unique synthesis of experimental data with the underlying physics, while allowing the latter to be less than perfect (i.e., approximate).[22-25] The present work is the first to incorporate experimental data by a minimal bias method into AIMD simulation, and specifically via the EDS approach.

We have intentionally chosen 300K water because it is a challenge for AIMD. However, our goal is to improve agreement with experiment for any system and DFT approximation, *not* just to create a better DFT variant for water. We show that a simple DFT approximation (BLYP) can achieve results consistent with state-of-the art approaches with the addition of a small amount of experimental data and at negligible increase in computational expense. The issue of over-structured AIMD water at ambient temperature has sometimes been addressed via indirect, arguably *ad hoc*, and/or empirical modifications, such as increasing the temperature or by adding a 2-body force to better model dispersion interactions. Increasing the level of theory beyond that of standard DFT approximations is also possible,[26-28] but a recent study showed that even computationally expensive Møller–Plesset perturbation theory does not reproduce all experimental properties of water.[29] Indeed, the remarkable success of a new "first principles" based potential for water, MB-Pol,[30-31] has shown that fitting the energy expansions through three-body terms to data obtained using very high level electronic structure calculations, as well as including additional many-body polarization effects in the final model, are required to accurately simulate the various properties of liquid water simultaneously at and near ambient conditions. This latter result does not bode well for the prospects that AIMD as a "first principles" (non-empirically modified) approach will be able to model liquid water accurately unless very high level electronic structure can somehow be utilized in the AIMD with sufficient computational speed and efficiency to ensure statistical convergence.

Minimally modifying the Hamiltonian of the AIMD simulation first requires a transformation of the experimental data into an analytical form that can be integrated into the equations of motion. X-ray scattering data of water from Skinner et al.[32] in the form of a radial distribution function (RDF) was transformed into simple scalars, i.e., the coordination number from the first solvation shell of water and three of its moments. This "coarse-graining" of the X-ray data into four scalars is done to make the bias as small as possible, yet still contain enough information to "repair" the over-structuring of water. Scalar properties extracted from other forms of available experimental data could also be straightforwardly incorporated for a general system. The justification for four scalars in this case was made based on testing one through five scalars on a computationally efficient classical empirical water model (Figure S1).

The Hamiltonian is minimally changed by adding a biasing term *linear* in the dimension to be modified.[20] The dimension is coordination number and its moments. Thus the following bias was added to the AIMD equations as additional potential energy for each oxygen atom:

$$V(r_i) = \sum_{k=0}^{3} \frac{\alpha_k}{\hat{f}_k} \sum_{j \neq i}^{N} r_{ij}{}^k [1 - u(r_0 - r_{ij})] \tag{1}$$

where $N$ is the number of neighbors of oxygen atoms, $r^k$ is the $k$th power of distance to coordinating atoms, $r_{ij}{}^k$ is the pairwise distance between two oxygen atoms raised to the $k$th power, $u(x)$ is a mollified unit step function (Equation S3), $r_0$ (2.5 Å) is approximately the location of the first peak of radial distribution function between oxygen atoms, $\alpha_k$ is found using the EDS method and chosen such that the coordination number and moments match those from the x-ray data, and $\hat{f}_k$ is the desired ensemble average for the $k$th scalar calculated from the experimental RDF according to equation S2. Note that $\hat{f}_k$ is not required in Eq. 1 but is there so that $\alpha_k$ is in units of energy. This biasing equation was found to be successful in Ref.[20] at modifying molecular structure using bias derived from RDFs.

Using this minimal bias built from the x-ray data, we conducted an EDS simulation of 128 water molecules in the constant NVT ensemble. All simulations were done with the Becke-Lee-Yang-Parr (BLYP) density functional, TZV2P basis set, Goedecker-Teter-Hutter (GTH) pseudopotentials, and in the CP2K Quickstep engine.[3] Three independent 20 ps BLYP-EDS simulations found the bias shown in Figure 1a, where it has been projected on the oxygen-oxygen distance coordinate. Note that the bias from 3 independent simulations overlaps within the thickness of the plot. Remarkably, even given the poor performance of the BLYP AIMD alone, the BLYP-EDS simulation provides a nearly perfect match between the experimental and AIMD RDF for oxygen-oxygen in pure water simulations at 300 K. Figure 1b shows this agreement from a 40 ps constant NVE simulation where the bias from Figure 1a was applied. The NVE ensemble was chosen so that self-diffusion data would also be valid and not affected by a thermostat. EDS does require simulation time to equilibrate to the experimental data (determine $\alpha_k$) and 2-5 ps were found sufficient to converge the bias within a few percent (Figure S2), although for all results here we used 20 ps. (See Ref. [20] for more discussion on convergence.) These results show that BLYP-EDS has a significantly improved oxygen-oxygen ($O_w$-$O_w$) RDF compared with the BLYP result and with no change in computational cost. A comparison of the added bias potential and the $O_w$-$O_w$ potential of mean force for the BLYP-EDS and BLYP simulations is shown in Figure S3 of the SI.

The result for the bias potential shown in Figure 1a and Figure S3 also provide physical insight into the flaws of the underlying BLYP DFT in the unbiased AIMD simulation of liquid water. The EDS method produces a bias potential in the form of an added repulsion to the AIMD simulation of a reasonably significant magnitude, and this bias potential works to overcome the unphysical degree of over-polarization and anomalous charge transfer at short range in the BLYP functional, which in turn leads, at least in part, to the over-structured and slowly diffusing AIMD water. Interestingly, this effect would seem to have little to do with the absence of longer range dispersion interactions in the BLYP DFT, which are often the target of "corrected" DFT.

The EDS bias is minimal in the relative entropy sense and thus only weakly perturbs other properties of the system, but it has been shown that the resulting biased ensemble moves closer to the distribution from a hypothetical perfect model agreeing with experiments[21] To verify this, we can check if other calculated properties have improved with respect to experiment, despite not having any explicit bias on those observables. The unbiased oxygen-hydrogen and hydrogen-hydrogen RDFs are shown in Figures 1c and 1d, respectively. We indeed observe significant improvements for the BLYP-EDS results versus those of the unbiased BLYP. As an example, the long-range over-structuring of the oxygen-hydrogen RDF in BLYP is eliminated with EDS.

Improving the static properties of water alone by EDS is not enough; the self-diffusion coefficient of water obtained from BLYP-level AIMD simulation can be too small by as much as two orders-of-magnitude compared with experiment, depending on the initial conditions in this glassy system. However, three independent 40 ps constant NVE BLYP-EDS simulations of water gave a self-diffusion coefficient for water of $0.06 \pm 0.02$ Å² / ps, which is the same order of magnitude as the experimental value (0.23 Å²/ ps).[33] This is a significant improvement over the original BLYP result computed in one simulation as 0.007 Å²/ ps (consistent with values 0.005 - 0.008 Å²/ ps appearing in past publications)[4]. We note that neither an increase in temperature nor any other change in simulation parameters was implemented aside from incorporating an experimentally determined bias potential via EDS. Nuclear quantum effects were not included in the simulation, which could potentially increase the self-diffusion coefficient.[34-35] Water simulations with small system-size have been shown to underestimate diffusivity coefficients (specifically by 24% for a 128 water molecule system, as used in this study).[36]

To summarize the work thus far, a bias on *only* the oxygen-oxygen coordination number and the first three moments of the $O_w$-$O_w$ RDF in AIMD water greatly improves the oxygen-oxygen (biased), oxygen-hydrogen (not biased), hydrogen-hydrogen (not biased) RDFs, as well as the diffusivity of water (not biased), with no increase in computational cost. It appears that the electronic structure of the system, even at the inaccurate BLYP DFT level, "responds" very favorably to an applied minimal bias of just a few experimental data points, and then provides an overall much better description of other, unbiased properties.

The EDS bias required to improve BLYP water is fairly high, around 0.5 kcal/mol (see Fig. 1a). Another test of EDS is to begin with a corrected DFT model that is already rather accurate and to then confirm that EDS still improves results but with less change. A natural choice of corrected DFT for such a study is BLYP with the D3 dispersion correction,[15] which is a DFT modification known to improve water structure and density. As described in the SI (see Fig. S4), the results are again significantly improved. The $O_w$-$O_w$ RDF gave the same near exact match with experiments (Figure S4b) with a smaller bias (Figure S4a). Interestingly, EDS and D3 provide complementary improvements and their combination gave a better $O_w$-$H_w$ RDF and $H_w$-$H_w$ RDF than either method independently as shown in Figure S4c-d. EDS is therefore not necessarily a replacement for empirical dispersion corrections and can instead be used supplement any DFT-based AIMD simulation methodology.

As another example, when applied to study the hydrated excess proton in water one of the biggest drawbacks of AIMD is that the simulation yields a ratio of proton to water diffusion that greatly disagrees with experiment. Excess protons migrate both through shuttling along hydrogen-bond connected water molecules (Grotthuss hopping)[12] and vehicular diffusion of a $H_3O^+$ hydronium ion. Experimental values give a 4:1 ratio of excess proton to water diffusivities,[33, 37] whereas AIMD simulations usually yield large ratios between 70:1 (BLYP) and 31:1 (BLYP-D2),[4, 38] mostly due to the slow underlying water diffusion. AIMD simulations using the BLYP functional can give a reasonably accurate excess proton diffusivity of 0.45 Å²/ps vs. 0.93 Å²/ps in experiment, but with an incorrect balance of transport mechanisms.[4] The reasonably fast proton transport in BLYP AIMD for the hydrated excess proton is likely related to the fact that the proton transfer barrier for BLYP is lower than it should be,[39] which compensates in part for the anomalously low underlying BLYP water self-diffusion in BLYP AIMD.

Importantly, the case of the hydrated excess proton also presents a test of the transferability of the EDS-AIMD-derived water model developed here, where *the same bias obtained for pure water* was applied to study a system with an excess proton. Three independent 60 ps constant NVE simulations of BLYP-EDS water gave an excess proton diffusion coefficient of 0.72 ± 0.42 Å²/ps, that is improved over the results from the original BLYP functional, but more importantly a significantly improved ratio of proton to water diffusion of 10:1 is obtained due to the underlying faster BLYP-EDS water diffusion. This change in mechanism of diffusion is partly due to the lifetime of hydrogen bonds decreasing (see Figure S5), which disrupts the structure required for a proton transfer to occur. The disruption of the hydrogen bonding can be seen more clearly in Figure S6, which shows the PMF of a hydrogen bond marginalized along the length of the bond and angle. EDS broadens the allowable geometry of hydrogen bonding. It is important to note that only coordination number between oxygen atoms in the water solvent was biased. Yet, that led to the proton diffusion, which is governed via chemical bonds breaking and forming, becoming significantly better with the *transferred* bias from neat water simulations.

The RDFs for the hydrated excess proton complex, which are shown in Figure S8 of the SI, further demonstrate the *minimal* biasing property of EDS. In general, the BLYP-EDS simulations appear to be the same or a small improvement over BLYP. (We note that care should be taken when comparing the AIMD results at pH = 0.43 to highly concentrated acid experiments (pH = – 0.75 pH).[4, 40])

The direct use of experimental data via EDS to modify a minimal set of water-water interactions in an AIMD simulation has been shown to lead to a significant improvement in the bulk water properties, including those that have not been biased. Moreover, the resulting EDS-AIMD model is transferable to the case of a hydrated excess proton in water where both structure and dynamics improve. The EDS method does not noticeably change the computational cost of AIMD simulations and requires only a few picoseconds of simulation time to determine a bias, which can then be subsequently applied in all future simulations. This improvement in the effective accuracy of the underlying DFT (BLYP for water in this case) without an increase in simulation time opens the door to more accurate AIMD simulations of a variety of systems. Future work will also demonstrate how the method can be applied to constant pressure AIMD simulations.

The idea of combining AIMD with experimental data is a fundamentally new approach, which promises to improve AIMD simulation models to match experiments for a general system without *ad hoc* modifications of the density functionals or simulation conditions. In general, as the amount of data generated from experiments increases, methods such as EDS represent a powerful way of utilizing this data through the lens of simulation. For example, EDS can be used immediately on other liquids. EDS is furthermore not restricted to experimental data; the data may come instead from higher-level quantum calculations. EDS can also be applied to semi-empirical and other non-DFT AIMD methods. Finally, we note that the EDS implementation used here is available as free and open-source software plug-in via PLUMED[41] for CP2K and can be easily adapted to other simulation engines.

**Supplementary Material**

A more detailed description of the implementation and outcomes of the EDS-AIMD algorithm, with additional results given for the EDS-AIMD simulation of BLYP+D3 water and the excess hydrated proton in water.

**Acknowledgements**

This research was supported by the Office of Naval Research (ONR grant N00014-15-1-2493). G.M.H. was supported as a Kadanoff-Rice postdoctoral scholar sponsored by the University of Chicago National Science Foundation MRSEC (grant DMR-1420709). The computational resources in this work were provided in part by a grant of computer time from the U.S. Department of Defense (DOD) High Performance Computing Modernization Program at the Engineer Research and Development Center (ERDC) and Navy DOD Supercomputing Resource Centers and in part by the University of Chicago Research Computing Center (RCC). This research also used resources of the Argonne Leadership Computing Facility, which is a DOE Office of Science User Facility supported under Contract DE-AC02-06CH11357. This research was also performed in part on the Stampede supercomputer at the Texas Advanced Computing Center (TACC), with resources provided by the Extreme Science and Engineering Discovery Environment (XSEDE), which is supported by National Science Foundation grant number ACI-1053575.

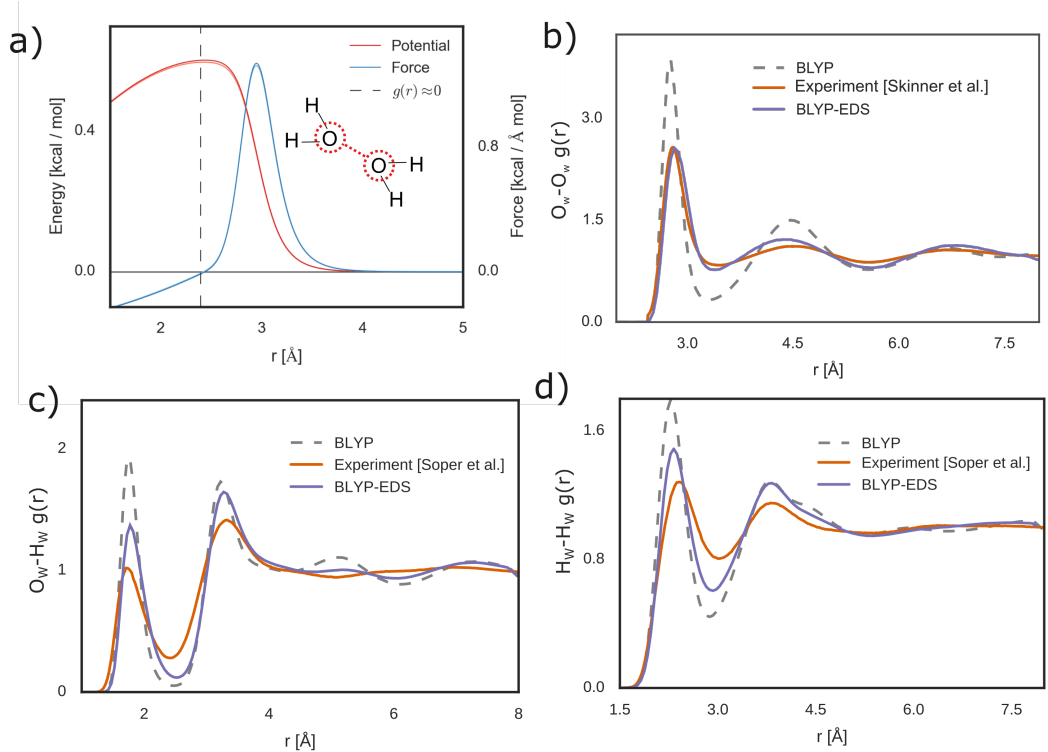

**Figure 1**: The resulting bias and fit to experimental data from an EDS-AIMD simulation. (a) The bias potential (red) and force (blue) applied, which is shown here as a projection onto the distance between water-oxygen atoms. The results from three replicates are shown but they overlap within the thickness of the lines. The vertical dashed line delineates the effectively zero-probability region in the $O_w$-$O_w$ radial distribution function, hence in practice only the bias to the right of that line is felt by the system. (b) The water oxygen-water oxygen ($O_w$-$O_w$) radial distribution functions from a constant NVE simulation between the EDS-AIMD using the equilibrated bias in (a) and experimental data from Skinner et al.[32] (c-d) The oxygen-hydrogen ($O_w$-$H_w$) and hydrogen-hydrogen ($H_w$-$H_w$) radial distribution functions from the same 40 ps EDS-AIMD simulation show substantial improvement despite not being biased directly. The experimental data for these (c-d) is from Soper and Benmore.[42]

# References

1. Car, R., and Parrinello, M., *Phys. Rev. Lett.* **55,** 2471 (1985).
2. Tuckerman, M., Laasonen, K., Sprik, M., and Parrinello, M., *J. Chem. Phys.* **103,** 150 (1995).
3. Vandevondele, J., Krack, M., Mohamed, F., Parrinello, M., Chassaing, T., and Hutter, J., *Comput. Phys. Commun.* **167,** 103 (2005).
4. Tse, Y. L. S., Knight, C., and Voth, G. A., *J. Chem. Phys.* **142,** 014104 (2015).
5. Izvekov, S., and Voth, G. A., *J. Chem. Phys.* **123,** 044505 (2005).
6. Todorova, T., Seitsonen, A. P., Hutter, J., Kuo, I. F. W., and Mundy, C. J., *J. Chem. Phys. B* **110,** 3685 (2005).
7. Baer, M. D., Mundy, C. J., Mcgrath, M. J., Kuo, I.-F. W., Siepmann, J. I., and Tobias, D. J., *J. Chem. Phys.* **135,** 124712 (2011).
8. Grossman, J. C., Schwegler, E., Draeger, E. W., Gygi, F., and Galli, G., *J. Chem. Phys.* **120,** 300 (2004).
9. Vandevondele, J., Mohamed, F., Krack, M., Hutter, J., Sprik, M., and Parrinello, M., *J. Chem. Phys.* **122,** 14515 (2005).
10. Yoo, S., Zeng, X. C., and Xantheas, S. S., *J. Chem. Phys.* **130,** 221102 (2009).
11. Yoo, S., and Xantheas, S. S., *J. Chem. Phys.* **134,** 121105 (2011).
12. Von Grotthuss, C. J. T., *Ann. Chim.* **58,** 54 (1806).
13. Agmon, N., *Chem. Phys. Lett.* **244,** 456 (1995).
14. Cukierman, S., *Bba-Bioenergetics* **1757,** 876 (2006).
15. Gillan, M. J., Alfe, D., and Michaelides, A., *J. Chem. Phys.* **144,** 130901 (2016).
16. Medvedev, M. G., Bushmarinov, I. S., Sun, J., Perdew, J. P., and Lyssenko, K. A., *Science* **355,** 49 (2017).
17. Distasio, R. A., Jr., Santra, B., Li, Z., Wu, X., and Car, R., *J. Chem. Phys.* **141,** 084502 (2014).
18. Fritz, M., Fernández-Serra, M., and Soler, J. M., *J. Chem. Phys.* **144,** 224101 (2016).
19. Mcgrath, M. J., Kuo, I. F. W., and Siepmann, J. I., *Phys. Chem. Chem. Phys.* **13,** 19943 (2011).
20. White, A. D., and Voth, G. A., *J. Chem. Theory Comput.* **10,** 3023 (2014).
21. Dannenhoffer-Lafage, T., White, A. D., and Voth, G. A., *J. Chem. Theory Comput.* **12,** 2144 (2016).
22. Pitera, J. W., and Chodera, J. D., *J. Chem. Theory Comput.* **8,** 3445 (2012).
23. Boomsma, W., Ferkinghoff-Borg, J., and Lindorff-Larsen, K., *Plos Comput. Biol.* **10,** e1003406 (2014).
24. Roux, B., and Weare, J., *J. Chem. Phys.* **138,** 084107 (2013).
25. White, A. D., Dama, J. F., and Voth, G. A., *J. Chem. Theory Comput.* **11,** 2451 (2015).
26. Guidon, M., Schiffmann, F., Hutter, J., and Vandevondele, J., *J. Chem. Phys.* **128,** 214104 (2008).
27. Kozuch, S., Gruzman, D., and Martin, J. M. L., *J. Phys. Chem. C* **114,** 20801 (2010).
28. Del Ben, M., Schonherr, M., Hutter, J., and Vandevondele, J., *J. Phys. Chem. Lett.* **4,** 3753 (2013).
29. Willow, S. Y., Zeng, X. C., Xantheas, S. S., Kim, K. S., and Hirata, S., *J. Phys. Chem. Lett.* **7,** 680 (2016).
30. Medders, G. R., Gotz, A. W., Morales, M. A., Bajaj, P., and Paesani, F., *J. Chem. Phys.* **143,** 104102 (2015).

31. Medders, G. R., Babin, V., and Paesani, F., *J. Chem. Theory Comput.* **10,** 2906 (2014).
32. Skinner, L. B., Huang, C. C., Schlesinger, D., Pettersson, L. G. M., Nilsson, A., and Benmore, C. J., *J. Chem. Phys.* **138,** 074506 (2013).
33. Krynicki, K., Green, C. D., and Sawyer, D. W., *Faraday Discuss.* **66,** 199 (1978).
34. Ceriotti, M., Cuny, J., Parrinello, M., and Manolopoulos, D. E., *Proc. Natl. Acad. Sci. USA* **110,** 15591 (2013).
35. Habershon, S., Markland, T. E., and Manolopoulos, D. E., *J. Chem. Phys.* **131,** 024501 (2009).
36. Yeh, I. C., and Hummer, G., *J. Phys. Chem. B* **108,** 15873 (2004).
37. Roberts, N. K., and Northey, H. L., *J. Chem. Soc. Farad. T. 1* **70,** 253 (1974).
38. Grimme, S., Antony, J., Ehrlich, S., and Krieg, H., *J. Chem. Phys.* **132,** 154104 (2010).
39. Ojamae, L., Shavitt, I., and Singer, S. J., *J. Chem. Phys.* **109,** 5547 (1998).
40. Botti, A., Bruni, F., Imberti, S., Ricci, M. A., and Soper, A. K., *J. Chem. Phys.* **121,** 7840 (2004).
41. Bonomi, M., Branduardi, D., Bussi, G., Camilloni, C., Provasi, D., Raiteri, P., Donadio, D., Marinelli, F., Pietrucci, F., Broglia, R. A., and Parrinello, M., *Comput. Phys. Commun.* **180,** 1961 (2009).
42. Soper, A. K., and Benmore, C. J., *Phys. Rev. Lett.* **101,** 065502 (2008).

# Supporting Information for

**Improved *Ab Initio* Molecular Dynamics by Minimally Biasing with Experimental Data**

Andrew D. White,[1,2] Chris Knight,[3] Glen M. Hocky,[1] Gregory A. Voth[1]

[1]*Department of Chemistry, James Franck Institute, and Institute for Biophysical Dynamics, The University of Chicago, 5735 S Ellis Ave., Chicago, Illinois 60637, USA*

[2]*Department of Chemical Engineering, University of Rochester, Rochester, New York 14627, USA*

[3]*Leadership Computing Facility, Argonne National Laboratory, 9700 South Cass Avenue, Argonne, Illinois 60439, USA*

AIMD simulations were carried out using Born-Oppenheimer molecular dynamics with the Quickstep module in the 2.5 branch of CP2K.[1] Density functional theory was used with the BLYP[2-3] functional and the Gaussian and Plane Waves (GPW) method.[4] Core electron states were treated with Goedecker-Teter-Hutter (GTH) pseudopotentials.[5] The orbital transformation method with a tolerance of $10^{-7}$ a.u. was applied to optimize the electron density at each MD step. The molecular dynamics integration timestep was 0.5 fs. Following Tse, Knight and Voth, the TZV2P basis sets were used with a 400 Ry plane wave cutoff.[6] The constant NVT simulations were done with a canonical sampling through velocity rescaling (CSVR) thermostat.[7] Experiment-directed simulation (EDS) were carried out using moments of the solvation shell number following White and Voth[8] as implemented in a modified version of PLUMED.[9] The bias adds an entropic-maximal (minimal change to system) pairwise potential energy to the oxygen-oxygen atoms. The equation for the bias on the *i*th oxygen atom is:

$$V(r_i) = \sum_{k=0}^{3} \frac{\alpha_k}{\hat{f}_k} \sum_{j \neq i}^{N} r_{ij}{}^k [1 - u(r_0 - r_{ij})] \tag{S1}$$

where $N$ is the number of neighbors of oxygen atoms, $r^k$ is the $k$th power of distance to coordinating atoms, $r_{ij}{}^k$ is the pairwise distance between two oxygen atoms raised to the $k$th power, $u(x)$ is the unit step function, $r_0$ (2.5 Å) is approximately the location of the first peak of radial distribution function between oxygen atoms, $\alpha_k$ is found using the EDS method, and $\hat{f}_k$ is the desired ensemble average for the $k$th scalar. Note that $\hat{f}_k$ is not necessary in Eq. S1 and is present to convert $\alpha_k$ into units of energy. The value of $\hat{f}_k$ is calculated from the x-ray radial distribution function according to:

$$\hat{f}_k = \rho \int_0^\infty dr\, [1 - u(r - r_0)] 4\pi r^{2+k} g(r) \tag{S2}$$

where $\rho$ is the number density and $g(r)$ is the radial distribution function.

While EDS is being used, $\alpha_k$ is a function of the coordination number, resulting in a many-body biasing force. Once the coefficients are found, the bias coefficients are fixed and this results in a potential that is pairwise. In order to have smooth forces, the following mollified unit-step function was used:

$$1 - u(r_0 - r) = \begin{cases} \dfrac{1 - \left(\frac{r - r_0}{w}\right)^6}{1 - \left(\frac{r - r_0}{w}\right)^{12}}, & r > r_0 \\ 1, & \text{otherwise} \end{cases} \tag{S3}$$

where $w$ is taken to be 0.7 Å and $r_0$ (2.5 Å) is approximately the location of the first peak of the radial distribution function between oxygen atoms. This choice was motivated by placing as much curvature of the mollified unit step into the region which needs to be biased, namely where over structuring of the radial distribution function is present. The EDS method finds the values of $\alpha_k$ that cause the simulation to match $\hat{f}_k$. The experiment derived oxygen-oxygen water radial distribution function from Skinner et al.[10] was used for EDS in this work. EDS requires a characteristic time-scale and energy scale to assist in finding the $\alpha_k$ coupling constants. Values of 25 fs and 1.0 Hartree were used, respectively. The $\alpha_i/\hat{f}_k$ coupling constants used for the hydrated proton simulations (shown in Figure 1a of the main text) were: -4.29x10$^{-5}$ Ha, 4.95x10$^{-03}$ Ha/Bohr, 1.09x10$^{\wedge -05}$ Ha/Bohr$^2$, and -2.37x10$^{-03}$ Ha/Bohr$^3$. The $\hat{f}_k$ set-points for all simulations were 2.88, 15.6 Bohr, 84.5 Bohr$^2$ and 463 Bohr$^3$, respectively

The bulk water AIMD simulations were composed of 128 water molecules in a 15.5118 Å cubic box and starting configurations were generated from equilibrated empirical SPC water simulations. The hydrated excess proton simulations were identical except for the substitution of a hydronium ion for a single water molecule. Due to the variability in proton diffusivity seen in past excess proton simulations,[6, 11-18] three independent simulations were conducted with different starting configurations. Starting configurations were generated with Packmol[19] and simulated using the classical molecular dynamics described below.

Diffusivity statistics were collected for 35-80 ps in the constant NVE ensemble with a 400 Ry basis-set cutoff after two stages of equilibration. The first-stage was a 2.5 ps constant NVT simulation with a CSVR time-constant of 0.1 ps and a 280 Ry basis-set cutoff. The second equilibration stage was a 10 ps NVT simulation with a CSVR time-constant of 1ps and a 400 Ry planewave cutoff. Unless otherwise stated, all property calculations were averaged over all replicates simulated.

Molecular dynamics simulations with a modified SPC/E water force-field were performed using LAMMPS[20] to determine the number of biasing moments to use.[21] To create an MD model which mimics the over-structured AIMD-BLYP model, the partial charges were changed from -0.82 to -0.94 for oxygen and from 0.41 to 0.47 for hydrogen. EDS was applied to this model following the same biasing parameters given above. The bias was equilibrated for 20 ps with a 0.1 ps Nose-Hoover thermostat. This was followed by a slower 1 ps time-constant Nose-Hoover thermostat for 10 ps and finally a 20 ps production to generate the radial distribution function statistics. The results of this testing are depicted in Fig. S1. Figure S2 depicts the development of the EDS bias in the EDS-AIMD simulation (see figure caption and main text), while Fig. S3 shows the final bias potential compared to the O-O potential of mean force for the EDS-AIMD and BLYP-level AIMD simulations. The potential of mean force is defined here as

$$W_{\text{O-O}}(r) = -k_B T \ln[g_{\text{O-O}}(r)] \quad . \tag{S4}$$

To generate the starting configurations for the AIMD models, a nonreactive hydronium force-field was used based on a recently parameterized reactive model.[22]

The diffusivities were calculated from mean-square displacement graphs with shifted times. The uncertainty in the diffusivities were calculated using ordinary least-squares regression, where the degrees of freedom were replaced with the effective number of samples as computed from the autocorrelation time and the diffusivity uncertainties were combined using error propagation.[23] A summary of the various AIMD and EDS-AIMD simulations in this work is given in Table S1. Diffusivities were not calculated when the simulation time was below 40 ps due to insufficient data.

**Table S1**: Summary of AIMD simulation systems described in this work.

| System | NVT Equil | NVE Prod | Replicate | D H$_2$O [Å$^2$/ ps] | D H$^+$ [Å$^2$/ ps] | N$_W$* | Avg T |
|---|---|---|---|---|---|---|---|
| 128-Water | 20 ps | 40 ps | 1 | 0.0087±0.0004 | - | 3.97 | 306.6 |
| EDS, 128-Water | 20 ps | 40 ps | 1 | 0.051±0.02 | - | 3.84 | 305.6 |
| EDS, 128-Water | 20 ps | 40 ps | 2 | 0.026±0.001 | - | 3.67 | 306.8 |
| EDS, 128-Water | 20 ps | 40 ps | 3 | 0.098±0.004 | - | 3.84 | 304.5 |
| EDS-D3, 128-Water | 16.5 ps | 20 ps | 1 | - | - | 3.88 | 299.9 |
| EDS-D3, 128-Water | 16.5 ps | 20 ps | 2 | - | - | 3.95 | 311.8 |
| EDS-D3, 128-Water | 16.5 ps | 15 ps | 3 | - | - | 3.90 | 310.0 |
| D3, 128-Water | 19 ps | 10 ps | 1 | - | - | 4.03 | 294.5 |
| D3, 128-Water | 19 ps | 10 ps | 2 | - | - | 4.06 | 301.0 |
| D3, 128-Water | 19 ps | 10 ps | 3 | - | - | 4.08 | 299.2 |
| EDS, 128-Water+Proton | 20 ps | 60 ps | 1 | 0.063±0.001 | 1.56±0.1 | 3.87 | 301.1 |
| EDS, 128-Water+Proton | 20 ps | 60 ps | 2 | 0.076±0.002 | 0.27±0.09 | 3.84 | 299.2 |
| EDS, 128-Water+Proton | 20 ps | 60 ps | 3 | 0.050±0.004 | 0.35±0.05 | 3.84 | 294.8 |
| 128 Water + Proton | 20 ps | 60 ps | 1 | 0.0075±0.0002 | 0.05±0.02 | 3.97 | 299.3 |

*This is the exact coordination number, where there is no smoothing of the unit step function. This is not directly biased, so there is some variability in how biasing the smooth coordination numbers affects the exact coordination number.

To test the effects of EDS on a BLYP-D3 DFT simulation, three replicate AIMD simulations of EDS-biased and unbiased BLYP-D3 simulations were conducted using the same protocol and simulation parameters as described earlier and in the main text, aside from the additional D3 dispersion correction. The $O_w$-$O_w$ RDF gave the same near exact match with experiments (Figure S4b) with a smaller bias (Figure S4a). Interestingly, EDS and D3 provide complementary improvements and their combination gave a better $O_w$-$H_w$ RDF and $H_w$-$H_w$ RDF than either method independently as shown in Figure S4c-d.

The data plotted in Figure S5 is the autocorrelation of a hydrogen bond where the ensemble average is over all possible hydrogen bonds over all times, $t$. The function $h(t)$ is an indicator function for the existence of a specific hydrogen bond at time $t$. Hydrogen bonds were defined using the usual geometric definition:[24] an oxygen-hydrogen distance of less than 2.45 Å, oxygen-oxygen distance of less than 3.5 Å and an oxygen-hydrogen distance unit vector, oxygen-oxygen distance unit vector dot product of > 150°. Justification that these criteria applies to hydrated excess proton complexes is shown in Figure S6. The dashed red vertical lines show the geometric definition above and the distance to EP axis is the distance to the excess proton as measured from the donor oxygen. Figure S7 shows that the geometric criteria defined above still divides a minimum in the PMF for hydrogen bonds near or involving the excess proton.

The RDFs for the hydrated excess proton, which are shown in Figure S8 below, further demonstrate the *minimal* biasing property of EDS. In general, the BLYP-EDS simulations appear to be the same or a small improvement over BLYP. It should also be noted from Figure S8d that BLYP-EDS only slightly raises the free energy barrier, i.e., the potential of mean force (PMF), along the special pair coordinate of the distorted Eigen cation, $H_9O_4^+$, while at the same time increasing the overall excess proton diffusion rate.

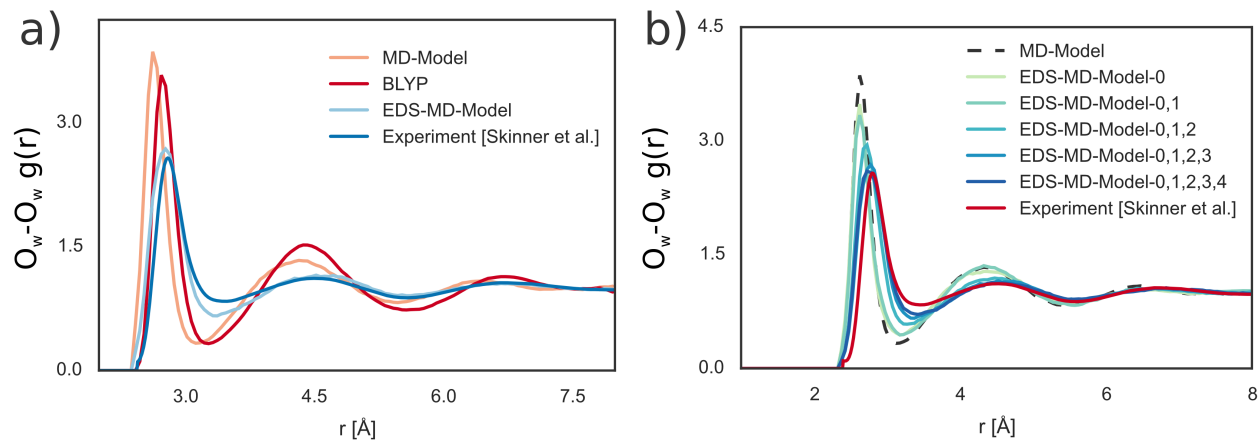

**Figure S1.** A molecular dynamics mimic (MD-Model) of DFT-AIMD was used to determine how best to bias AIMD with EDS method. Panel a shows the approximate agreement between the MD-Model and DFT-AIMD water oxygen-water oxygen ($O_w$-$O_w$) radial distribution functions (RDFs). Panel b, the line labeled EDS-MD-Model-0 shows how biasing first solvation shell number of $O_w$-$O_w$ to match the experimental value improves. Adding moments of solvation shell number leads to increasingly better fit. Four solvation-number moments was determined as sufficient based on Panel b and used for the DFT-AIMD simulations.

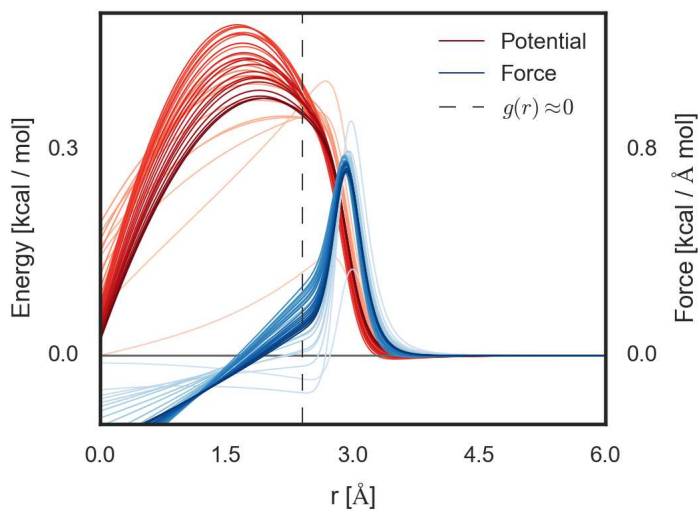

**Figure S2.** The development of the EDS bias over the first 5 ps of an EDS-AIMD simulation. The lighter lines are earlier time and darker lines are lighter in time. The darkest line is the bias after 5 ps. Note that the rapidly changing lines left of the vertical dashed lines are in zero-

probability regions. Only potential right of the vertical dashed lines is actually felt by the system (see Fig. 1A of the main text).

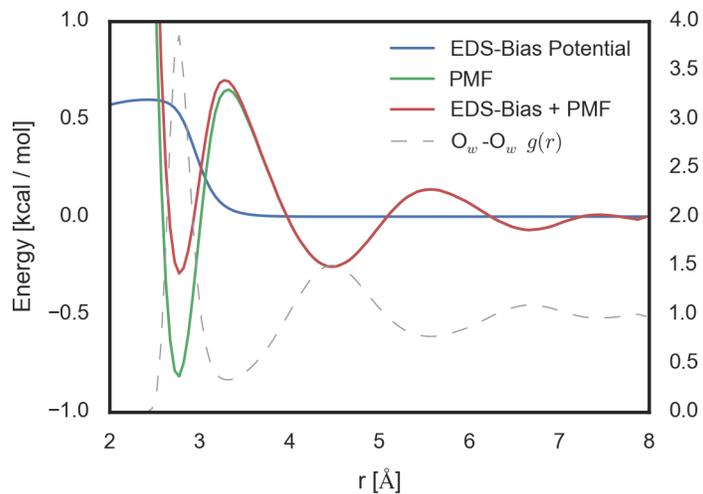

**Figure S3.** A comparison of the EDS biasing potential and the potential of mean force from Equation S3. The $O_w$-$O_w$ RDF and EDS bias are averaged from the pure water constant NVE simulations. The change in PMF is localized to the first peak in the RDF and reduces the attractive potential there by 0.5 kcal /mol. The right y-axis is g(r).

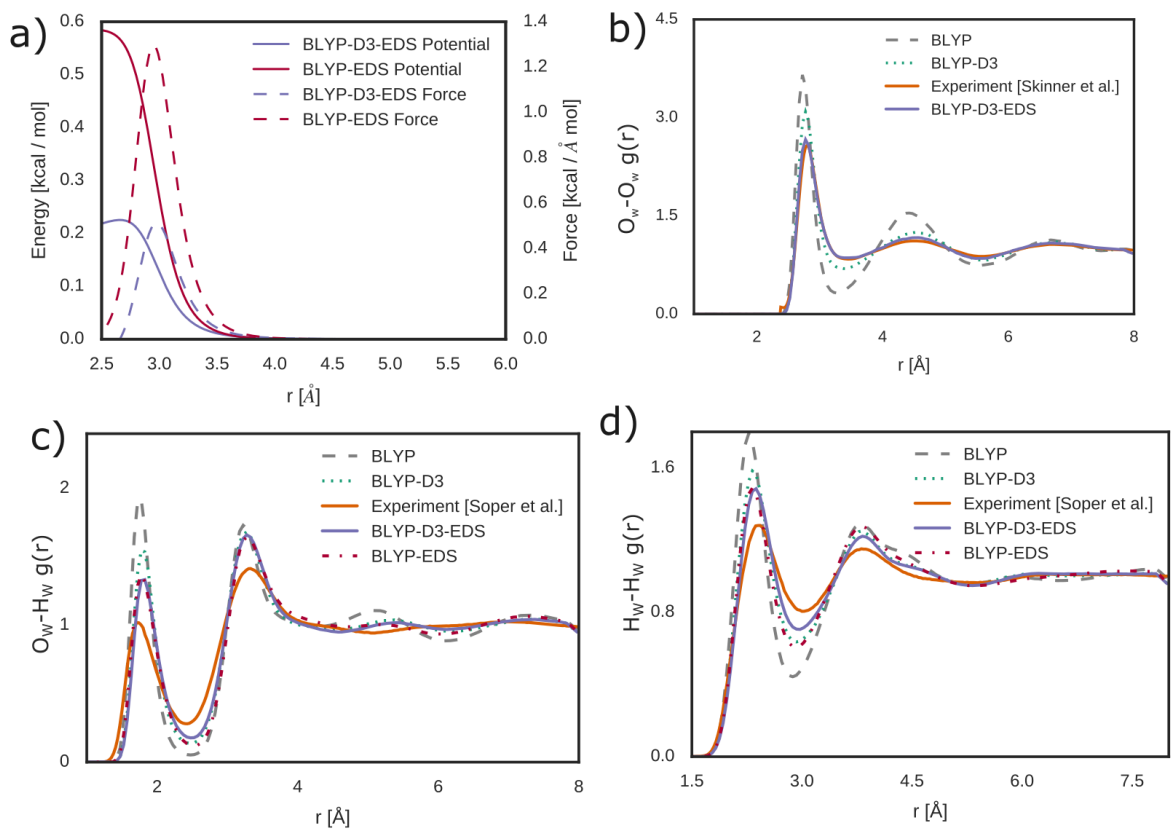

**Figure S4.** The resulting bias and fit to experimental data from a EDS-D3-AIMD simulations. (a) The bias potential and force for EDS-AIMD (red) and EDS-D3-AIMD (blue), which is shown here as a projection onto the distance between water-oxygen atoms. This shows less bias is necessary to converge when using the D3 dispersion correction. (b) The water oxygen-water oxygen ($O_w$-$O_w$) radial distribution functions from a constant NVE simulation using the equilibrated bias in (a) and experimental data from Skinner et al.[10] (c-d) The oxygen-hydrogen ($O_w$-$H_w$) and hydrogen-hydrogen ($H_w$-$H_w$) radial distribution functions from the same data as (a) simulation show substantial improvement despite not being biased directly. The experimental data for these (c-d) is from Ref.[25]. EDS and D3 combined shows better performance than either independently in these two radial distribution functions.

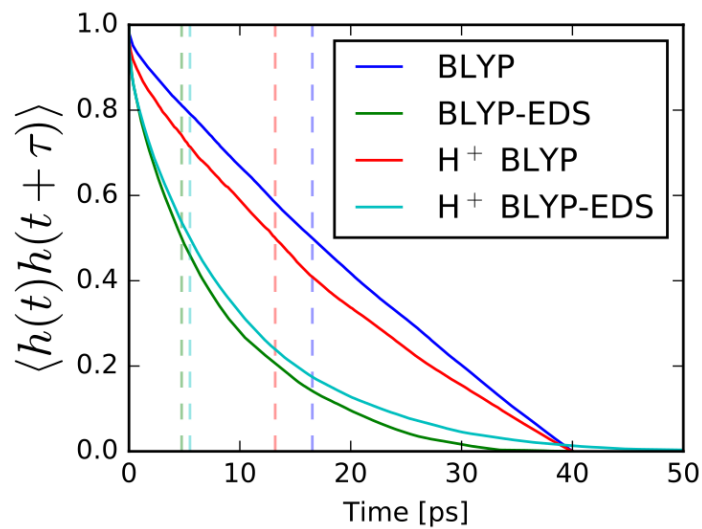

**Figure S5.** A comparison of hydrogen bonding autocorrelation function. The vertical line is the median hydrogen bonding lifetime. As seen through this and other measures, EDS breaks some of the slower hydrogen bonding structure seen in both BLYP and H$^+$BLYP. The reason for the non-exponential behavior of the unbiased simulations is that many h-bonds are long-lived and we truncated the BLYP hydronium simulation to be 40 ps to compare directly with the neat water simulation.

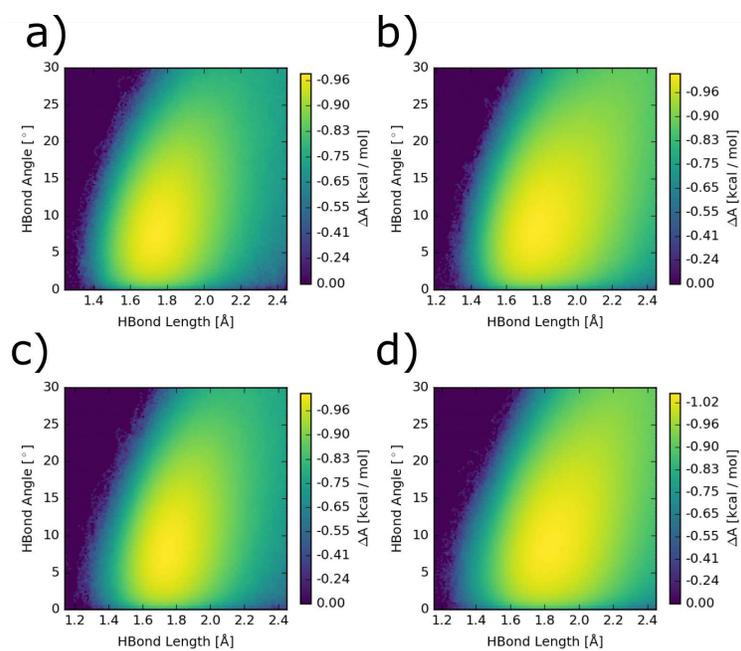

**Figure S6.** The potential of mean force for the hydrogen bonding structure in the BLYP, BLYP-EDS systems. a) and b) are water simulations, c) and d) have an excess proton (hydronium). The first column is BLYP and the second column is BLYP-EDS. EDS broadens the PMF of the hydrogen bonding, allowing more freedom in their formation.

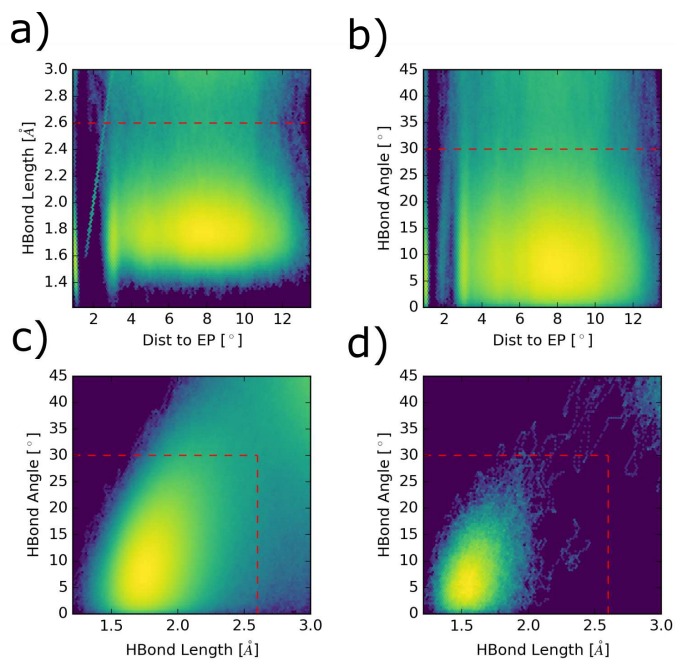

**Figure S7.** The potential of mean force for the hydrogen bonding structure in the BLYP excess proton simulations. The dashed red lines show the geometric hydrogen bonding criteria defined in the SI text. Panels a and b are plots of hydrogen bond length and angle PMF plotted as a function of distance to excess proton for all hydrogen bonds. Panel c is the hydrogen bonding criteria PMF for all hydrogen bonds and d is the same but for hydrogen bonds involving an excess proton only.

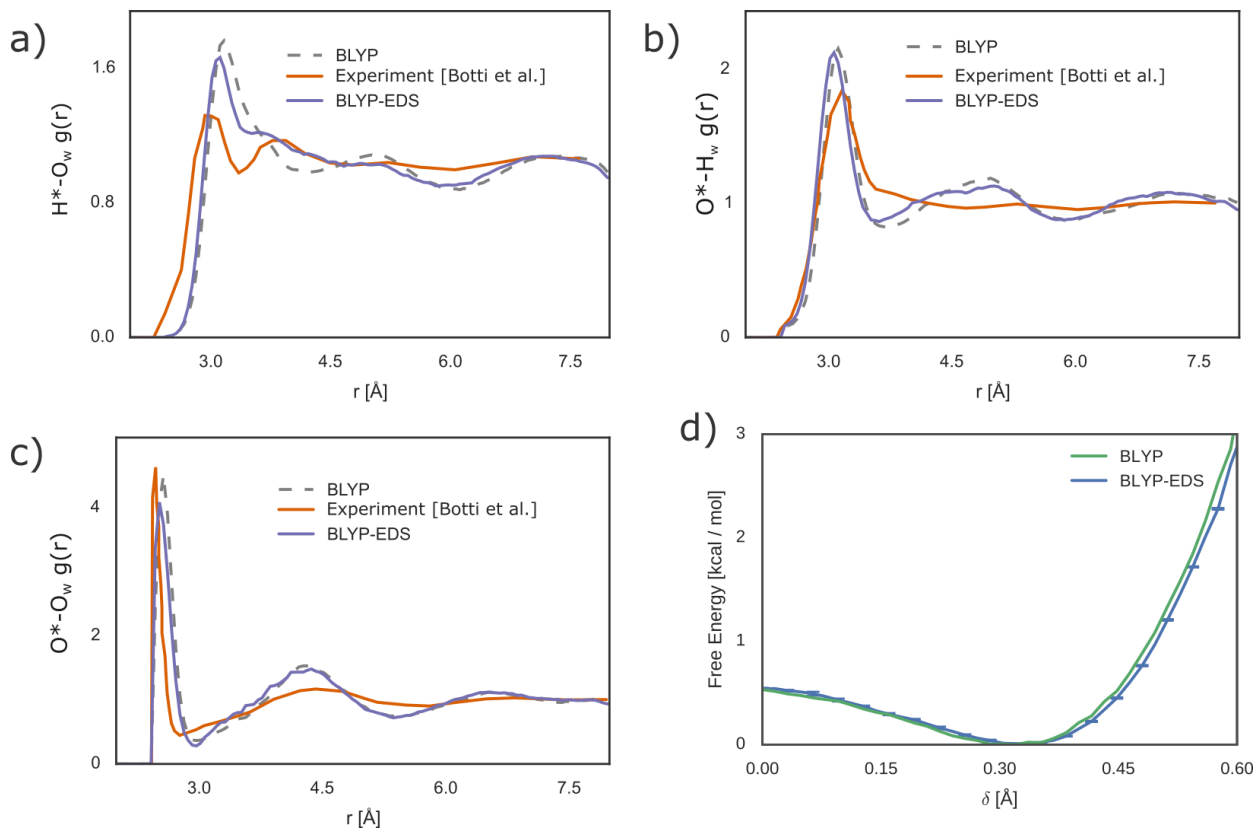

**Figure S8**: Equilibrium properties of a hydrated excess proton from 3 independent 60 ps NVE EDS-AIMD simulations *using the bias obtained for pure water* (Figure 1a of the main text). Panels a, b, and c show the radial distribution functions between the most hydronium-like ion and water. Panel d shows the potential of mean force (PMF) along the special pair coordinate of the distorted Eigen cation $H_9O_4^+$. The error bars are standard deviations across the 3 independent simulations. Some care should be taken directly comparing with experimental RDFs in panels a – c since they are from a pH -0.75 acid solution whereas the simulation is at pH 0.43. Note that since EDS is a minimal biasing method, these unbiased RDFs (panels a,b) should remain close to the corresponding unbiased simulation results.

**Supplemental References**

1.  Vandevondele, J., Krack, M., Mohamed, F., Parrinello, M., Chassaing, T., and Hutter, J., *Comput. Phys. Commun.* **167,** 103 (2005).
2.  Becke, A. D., *Phys. Rev. A* **38,** 3098 (1988).
3.  Lee, C. T., Yang, W. T., and Parr, R. G., *Phys. Rev. B* **37,** 785 (1988).
4.  Lippert, G., Hutter, J., and Parrinello, M., *Theor. Chem. Acc.* **103,** 124 (1999).
5.  Goedecker, S., Teter, M., and Hutter, J., *Phys. Rev. B* **54,** 1703 (1996).
6.  Tse, Y. L. S., Knight, C., and Voth, G. A., *J. Chem. Phys.* **142,** 014104 (2015).
7.  Bussi, G., Donadio, D., and Parrinello, M., *J. Chem. Phys.* **126,** 014101 (2007).
8.  White, A. D., and Voth, G. A., *J. Chem. Theory Comput.* **10,** 3023 (2014).
9.  Bonomi, M., Branduardi, D., Bussi, G., Camilloni, C., Provasi, D., Raiteri, P., Donadio, D., Marinelli, F., Pietrucci, F., Broglia, R. A., and Parrinello, M., *Comput. Phys. Commun.* **180,** 1961 (2009).
10. Skinner, L. B., Huang, C. C., Schlesinger, D., Pettersson, L. G. M., Nilsson, A., and Benmore, C. J., *J. Chem. Phys.* **138,** 074506 (2013).
11. Tuckerman, M., Laasonen, K., Sprik, M., and Parrinello, M., *J. Phys. Chem.* **99,** 5749 (1995).
12. Tuckerman, M., Laasonen, K., Sprik, M., and Parrinello, M., *J. Chem. Phys.* **103,** 150 (1995).
13. Lobaugh, J., and Voth, G. A., *J. Chem. Phys.* **104,** 2056 (1996).
14. Schmitt, U. W., and Voth, G. A., *J. Phys. Chem. B* **102,** 5547 (1998).
15. Lapid, H., Agmon, N., Petersen, M. K., and Voth, G. A., *J. Chem. Phys.* **122,** 014506 (2005).
16. Markovitch, O., Chen, H., Izvekov, S., Paesani, F., Voth, G. A., and Agmon, N., *J. Phys. Chem. B* **112,** 9456 (2008).
17. Hassanali, A., Giberti, F., Cuny, J., Kuhne, T. D., and Parrinello, M., *Proc. Natl. Acad. Sci. U. S. A.* **110,** 13723 (2013).
18. Hassanali, A. A., Giberti, F., Sosso, G. C., and Parrinello, M., *Chem. Phys. Lett.* **599,** (2014).
19. Martinez, L., Andrade, R., Birgin, E. G., and Martinez, J. M., *J. Comput. Chem.* **30,** 133 (2009).
20. Plimpton, S., *J. Comput. Phys.* **117,** 1 (1995).

21. Berendsen, H. J. C., Grigera, J. R., and Straatsma, T. P., *J. Phys. Chem.* **91,** 6269 (1987).
22. Tse, Y.-L. S., Chen, C., Lindberg, G. E., Kumar, R., and Voth, G. A., *J. Am. Chem. Soc.* **137,** 12610 (2015).
23. Müller-Krumbhaar, H., and Binder, K., *J. Stat. Phys.* **8,** 1 (1973).
24. Luzar, A., and Chandler, D., *Phys. Rev. Lett.* **76,** 928 (1996).
25. Soper, A. K., and Benmore, C. J., *Phys. Rev. Lett.* **101,** 065502 (2008).